\documentclass[12pt]{article}
\usepackage{graphics, amsfonts, amsmath, mathrsfs, amssymb, overcite}
\def\preprint
#1{\thispagestyle{empty}~\newline\vspace*{-22.65mm}
\begin{flushright}
\begin{tabular}{l} #1
\end{tabular}
\end{flushright}
\vspace{1cm}}

\begin{document}
\title{\bf \LARGE Ultrastatic spacetimes}
\author{Sebastiano Sonego\thanks{\tt sebastiano.sonego@uniud.it}
\\[4mm]
{\small\it Dipartimento di Fisica, Universit\`a di Udine}\\ 
{\small\it Via delle Scienze 208, 33100 Udine, Italy}}
\date{}
\maketitle
\begin{abstract}
Several calculations in conformally static spacetimes rely on the introduction of an ultrastatic background.  I describe the general properties of ultrastatic spacetimes, and then focus on the problem of whether a given spacetime can be ultrastatic, or conformally ultrastatic, in more than one way.  I show that the first possibility arises iff the spacetime contains regions that are products with a Minkowskian factor, and that the second arises iff it contains regions whose spatial sections are conformal to a product space.\\

\noindent PACS: 02.40.Ky; 04.20.Cv; 04.90.+e \\
Keywords: Ultrastatic spacetimes; Killing vector fields; conformal Killing vector fields; spacetime symmetries
\end{abstract}
\thispagestyle{empty}
\def\ee{{\mathrm e}}
\def\dd{{\mathrm d}}
\def\g{\mbox{\sl g}}
\def\tg{\widetilde{\mbox{\sl g}}}
\def\SIZE{1.00}
\newcommand{\Dd}{{\rm D}}
\newcommand{\nab}{\nabla\!}
\newcommand{\tnab}{\widetilde{\nabla}\!}

\newpage


\section{Introduction}
\label{sec1}
\setcounter{equation}{0}

A spacetime is ultrastatic if it admits an atlas of charts in which the metric tensor takes the form~\cite{fn1}
\begin{equation}
\g=-\dd t^2+\g_{ij}\,\dd x^i\dd x^j\;,
\label{steph}
\end{equation}
where the coefficients $\g_{ij}$ do not depend on the coordinate $t$. Ultrastatic spacetimes are interesting for several reasons. First, they possess the property that the possible paths of light rays coincide with the geodesics of the spatial metric $\g_{ij}\,\dd x^i\dd x^j$.  This follows immediately from Fermat's principle, which requires that $\int\dd t$, hence also 
\[ \int\left(\g_{ij}\,\dd x^i\dd x^j\right)^{1/2}\;,\] 
be extremal for light propagation.  Second, in these spacetimes there are no gravitational forces in the following sense.  If one computes the connection coefficients ${\Gamma^a}_{bc}$ for the metric~\eqref{steph}, one finds that only the components ${\Gamma^i}_{jk}$ can be nonzero. Since gravitational forces are commonly associated with the ${\Gamma^i}_{00}$ and ${\Gamma^i}_{0j}$ components, only ``inertial'' forces of the type ${\Gamma^i}_{jk}\,\dot{x}^j\dot{x}^k$ act on a freely falling particle (here, a dot denotes the derivative with respect to a suitable parameter --- {\em e.g.\/}, proper time --- along the particle world line)~\cite{fn2}.  Third, ultrastatic spacetimes are the only ones that admit a timelike vector field $\eta^a$ that is covariantly constant, {\em i.e.\/}, such that $\nab_a \eta^b=0$~\cite{stephani}.  Since the condition $\nab_a\eta^b=0$ expresses the fact that $\eta^a$ does not accelerate, rotate, or deform, such a vector field is an appropriate extension of the notion of an inertial frame in Minkowski spacetime. Related to this property, is the fact that in conformally static spacetimes (which, unlike the ultrastatic ones, include several cases of great physical relevance) one can simplify the description of many processes by reformulating them in terms of the conformally related ultrastatic background.  This procedure has led to a unified explanation of a large number of otherwise puzzling effects~\cite{???}, as well as to remarkable formal simplifications~\cite{optmet}.  Finally, because of their very simple structure, which makes them just curved {\em space\/} generalisations of Minkowski spacetime, ultrastatic spacetimes are also useful for educational purposes.

The goal of this article is to provide a reference that summarises the main properties of ultrastatic spacetimes.  Although some of the results are elementary, to the author's knowledge they are not reported in the existing literature.  In the next section, a chart-independent definition will be given, from which the coordinate representation~\eqref{steph} of the metric is derived. Furthermore, it will be shown explicitly that all the differential geometrical features of an ultrastatic spacetime, in particular its curvature tensors, are completely encoded in the spatial metric.  Section~\ref{sec3} is devoted to the problem of finding the class of coordinate transformations that preserve the form~\eqref{steph} of the metric.  Similarly, in section~\ref{sec4} I investigate when two ultrastatic metrics are nontrivially conformal to each other.  A brief summary of the results obtained is given in section~\ref{sec5}.


\section{Definition and general properties}
\label{sec2}
\setcounter{equation}{0}

A spacetime $(\mathscr{M},\g)$ is called {\em ultrastatic\/} iff it is static --- {\em i.e.\/}, it possesses a hypersurface-orthogonal timelike Killing vector field $\eta^a$ --- and, in addition, $\eta^a$ has constant norm~\cite{ultra}.  Without loss of generality, one can suppose that $\eta^a$ has unit norm, {\em i.e.\/}, $\g_{ab}\,\eta^a\eta^b=-1$.  This is achieved merely through a uniform rescaling of the Killing parameter.

Alternatively, ultrastatic spacetimes can be characterised requiring that there exist a covariantly constant timelike vector field.  This follows immediately from the definition and a lemma: $\eta^a$ is a hypersurface-orthogonal Killing vector field with constant norm iff it is covariantly constant.  To prove this statement, write the condition for hypersurface-orthogonality, $\eta_{[a}\nab_b\eta_{c]}=0$, in the form
\begin{equation}
\eta_a\nab_b\eta_c+\eta_b\nab_c\eta_a+\eta_c\nab_a\eta_b=0\;,
\label{ho}
\end{equation}
where Killing's equation $\nab_a\eta_b+\nab_b\eta_a=0$ has been used.  Contracting equation~\eqref{ho} by $\eta^a\eta^b$, and imposing the condition that $\eta^a$ has constant norm in the form $\eta^a\nab_b\eta_a=0$, one finds immediately that $\eta^a$ is geodesic.  Finally, contracting equation~\eqref{ho} by $\eta^c$, one obtains the desired result, $\nab_a\eta_b=0$.  (For an alternative proof, one can use equation~(C.3.12) in reference~[8].)  The converse implication, that if $\eta^a$ is covariantly constant then it is a hypersurface-orthogonal Killing vector field with constant norm, is trivial.

The condition $\nab_a\eta_b=0$ implies that the one-form $\eta_a$ is closed, so locally one can introduce a function $t$ on $\mathscr{M}$ such that $\eta_a=-\nab_a t$.  Then, one can write
\begin{equation}
\g_{ab}=-\nab_at\,\nab_bt+h_{ab}\;,
\label{nabtnabt}
\end{equation}
where $h_{ab}$ is transverse to $\eta^a$ ({\em i.e.\/}, $\eta^ah_{ab}=0$)  and $\pounds_\eta\, h_{ab}=\pounds_\eta\,\g_{ab}=0$.  Introducing, on the $t=\mbox{const}$ hypersurfaces, coordinates $x^i$ that are constant on the integral curves of $\eta^a$, and using $t$ as the coordinate $x^0$, it is easy to see that $\eta^a=\delta^a_0$, so $t$ is also the Killing parameter.  Then, $h_{ab}\,\dd x^a\dd x^b=\g_{ij}\,\dd x^i\dd x^j$ and equation~\eqref{nabtnabt} leads to the form~\eqref{steph} for the spacetime metric.  Of course, this is just the particular case of a static metric (see, {\em e.g.\/}, reference~[8], p.~119) with $\g_{00}=\g_{ab}\,\eta^a\eta^b=-1$.  One can consider the quotient space $\mathscr{S}$ of $\mathscr{M}$ with respect to the equivalence relation defined by the isometry generated by $\eta^a$.  Clearly, $\mathscr{S}$ has a structure of three-dimensional Riemannian manifold with metric $h=h_{ab}\,\dd x^a\dd x^b$, and the hypersurfaces $t=\mbox{const}$ are all locally isometric to $(\mathscr{S},h)$.  Since $\eta^a$ is the unit vector orthogonal to these hypersurfaces, $(\mathscr{S},h)$ can be regarded as the ``rest space'' of the observers with four-velocity $\eta^a$.  This is an invariant notion based on the symmetries of $(\mathscr{M},\g)$, so space is unique if $\eta^a$ is.  

The differential geometrical properties of ultrastatic spacetimes are completely determined by the spatial metric $h$ --- not surprisingly, considering equations~\eqref{steph} and~\eqref{nabtnabt} and the fact that, in the coordinates $t$ and $x^i$, ${\Gamma^0}_{ab}={\Gamma^a}_{b0}={\Gamma^i}_{jk,0}=0$~\cite{stephani}.  In particular, the only components of the Riemann tensor ${R_{abc}}^d$ of $(\mathscr{M},\g)$ that do not vanish identically, coincide with those of the Riemann tensor $^{(3)}\!{R_{abc}}^d$ of $(\mathscr{S},h)$.  In order to prove this, let us first notice that $(\mathscr{S},h)$ has vanishing extrinsic curvature into $(\mathscr{M},\g)$, as it follows immediately from the property $2K_{ab}=\pounds_\eta\, h_{ab}=0$ (alternatively, since $\eta^a$ is covariantly constant, $K_{ab}={h_a}^c\,\nab_c\eta_b=0$).  As a consequence, one of the Gauss-Codazzi equations implies that the Riemann tensor of $(\mathscr{S},h)$ is simply given by $^{(3)}\!{R_{abc}}^d={h_a}^{a'}{h_b}^{b'}{h_c}^{c'}{h^d}_{d'}{R_{a'b'c'}}^{d'}$.  Next, notice that ${R_{abc}}^0={R_{abc}}^d\,\eta_d=\nab_a\nab_b\eta_c- \nab_b\nab_a\eta_c=0$.  Together with the symmetry properties of the Riemann tensor, this implies that only the purely spatial components of ${R_{abc}}^d$ can be nonzero (equivalently, any  contraction of ${R_{abc}}^d$ with $\eta^a$ must vanish). Then we have simply ${R_{abc}}^d={h_a}^{a'}{h_b}^{b'}{h_c}^{c'}{h^d}_{d'}{R_{a'b'c'}}^{d'}= {^{(3)}\!{R_{abc}}^d}$.  Since this implies $R_{ab}={}^{(3)}\!R_{ab}$, the second Gauss-Codazzi equation is automatically satisfied.

{From} these expressions for the curvature in an ultrastatic spacetime, one can write the Einstein tensor as 
\begin{equation}
G_{ab}=\frac{1}{2}\,{}^{(3)}\!R\,\nab_a t\,\nab_b t +{}^{(3)}\!R_{ab}-\frac{1}{2}\,{}^{(3)}\!R\,h_{ab} =\frac{1}{2}\,{}^{(3)}\!R\,\nab_a t\,\nab_b t+{}^{(3)}G_{ab}\;,
\end{equation}
which obviously decomposes into a part ``parallel'' to $\eta^a$ and one transverse to it.  Replacing this into Einstein's equations $G_{ab}+\Lambda\, \g_{ab}=\kappa\, T_{ab}$, one obtains:
\begin{equation}
T_{ab}\,\eta^a {h^b}_c=0\;;
\label{flux}
\end{equation}
\begin{equation}
^{(3)}\!R=\kappa\left(3\,T_{ab}\,\eta^a\eta^b+T_{ab}\,h^{ab}\right)\;;
\label{R}
\end{equation}
\begin{equation}
\Lambda=\frac{1}{2}\,\kappa\left(T_{ab}\,\eta^a\eta^b+T_{ab}\,h^{ab}\right)\;.
\label{energy}
\end{equation}
Note that both the weak and the strong energy conditions imply $^{(3)}\!R\geq 0$.  Similarly, the strong energy condition implies $\Lambda\geq 0$.  Hence, an ultrastatic spacetime with negative spatial curvature and/or a negative cosmological constant requires ``exotic matter''.  In any case, the energy flux in the frame $\eta^a$ vanishes, according to equation~\eqref{flux}.

Finally, it is useful to write the expression for the Weyl tensor $C_{abcd}$ in an ultrastatic spacetime.  From equation~\eqref{nabtnabt} and the definition of $C_{abcd}$ one finds
\begin{equation}
C_{abcd}=\nab_a t\,\nab_{[c}t\, {}^{(3)}\!R_{d]b}- \nab_b t\,\nab_{[c}t\, {}^{(3)}\!R_{d]a}- \frac{1}{3}\,{}^{(3)}\!R \left(\nab_a t\,\nab_{[c}t\, h_{d]b}+ h_{a[c}\,\nab_{d]}t\,\nab_b t\right)\;,
\label{weyl}
\end{equation}
where the fact that the Weyl tensor of any three-dimensional manifold --- hence, in particular, of $(\mathscr{S},h)$ --- is equal to zero has been used.


\section{Ultrastatic transformations}
\label{sec3}
\setcounter{equation}{0}

When discussing invariance properties in general relativity, one can take an ``active'' or a ``passive'' view of the transformations that preserve a given property~\cite{wald}.  In the first case, a spacetime is regarded as the equivalence class of all pairs $(\mathscr{M},\g)$ that are related by diffeomorphisms.  One is then interested in the set of the particular diffeomorphism $\varphi$ such that both $\g$ and $\varphi_{\ast}\,\g$ possess the property one is interested in.  On the other hand, in the ``passive'' viewpoint, one's attention is focussed on a particular pair $(\mathscr{M},\g)$, and invariance is characterised by the transformations between charts on $\mathscr{M}$ that preserve some particular feature of the metric coefficients. These two descriptions are mathematically equivalent (although conceptually rather different), and one can use indifferently one or the other.  In the following, I shall focus on the passive view.

All the properties listed in the previous section are particularly evident in the coordinates $t$ and $x^i$ used in equation~\eqref{steph}, in which $\eta^a =\delta^a_0$. Such coordinates are therefore privileged for describing ultrastatic spacetimes:  They play a role similar to the one played by Lorentzian coordinates in Minkowski spacetime, being adapted to the observers who describe physical phenomena in the simplest way.  For this reason, I shall refer to the coordinate representation~\eqref{steph} of the metric as the {\em canonical form\/}, and to coordinates $t$ and $x^i$ in which equation~\eqref{steph} holds as {\em canonical coordinates\/}.  It is then interesting to find the set of transformations that map canonical coordinates to canonical coordinates, hereafter called {\em ultrastatic transformations\/} of $(\mathscr{M},\g)$.  This is the analogue, for ultrastatic spacetimes, of the Poincar\'e group in Minkowski spacetime, whose elements map Lorentzian coordinates to Lorentzian coordinates. More explicitly, let $x^a$ be canonical coordinates on the ultrastatic spacetime $(\mathscr{M},\g)$; a coordinate transformation $x^a\to \bar{x}^a$ on $\mathscr{M}$ is ultrastatic iff
\begin{equation}
-\dd t^2+\g_{ij}\,\dd x^i\dd x^j=-\dd\bar{t}^2+\bar{\g}_{ij}\, \dd\bar{x}^i\dd\bar{x}^j\;,
\label{coord}
\end{equation}
with $\partial\bar{\g}_{ij}/\partial\bar{t}=0$.  Obviously, transformations of the spatial coordinates alone that do not involve the time coordinate are of this kind, but they are not very interesting, since they only represent time-independent coordinate transformations on $\mathscr{S}$.  By excluding them, together with time translations, one remains with what can be called {\em proper ultrastatic transformations\/}.

Let us look for the generators of ultrastatic transformations, {\em i.e.\/}, for the vector field $\xi^a$ that appears in the most general infinitesimal transformation between canonical coordinates,
\begin{equation}
x^a\to \bar{x}^a=x^a+\varepsilon\,\xi^a(x)\;,
\label{transf}
\end{equation}
where $x^0\equiv t$, $\bar{x}^0\equiv \bar{t}$, and $\varepsilon$ is a parameter~\cite{discrete}.  The transformation~\eqref{transf} implies a change $\g_{ab}(x)\to\bar{\g}_{ab}(\bar{x})$ in the metric coefficients, with
\begin{equation}
\bar{\g}_{ab}(\bar{x})=\g_{ab}(\bar{x})-\varepsilon\,\pounds_\xi\,\g_{ab}(x)+{\cal O}(\varepsilon^2)\;.
\label{ggbar}
\end{equation}
In order for $\xi^a$ to generate an ultrastatic transformation, we must have $\bar{\g}_{00}=-1$, $\bar{\g}_{0i}=0$, and $\partial\bar{\g}_{ij}/\partial\bar{t}=0$.  Using the expression $\pounds_\xi\,\g_{ab}=\nab_a\,\xi_b+\nab_b\,\xi_a$, the first condition gives $\nab_0\xi_0=0$, {\em i.e.\/}, $\partial\xi^0/\partial t=0$, which implies $\xi^0=\alpha$, with $\alpha$ a function of the spatial coordinates $x^i$ only.  The second condition, $\nab_0\xi_i+\nab_i\xi_0=0$, becomes then $\partial\xi_i/\partial t=\partial\alpha/\partial x^i$, which gives
\begin{equation}
\xi^i=t\,\g^{ij}\,\partial_j\alpha+\beta^i\;,
\label{epsilon}
\end{equation}
where the $\beta^i$ do not depend on $t$.  Finally, let us take the derivative with respect to $t$ of equation~\eqref{ggbar} for $a=i$, $b=j$.  Expanding $\g_{ij}(\bar{x})$ around $x$, using the relation
\begin{equation}
\frac{\partial}{\partial t}=\frac{\partial}{\partial\bar{t}}+ \varepsilon\,\frac{\partial\xi^k}{\partial t}\,\frac{\partial}{\partial\bar{x}^k}\;,
\label{ttbar}
\end{equation}
and keeping only first order terms in $\varepsilon$, we find
\begin{equation}
\nab_i\left(\partial\xi_j/\partial t\right)+\nab_j\left(\partial\xi_i/\partial t\right)=0\;.
\label{Dxi}
\end{equation}
Using equation~\eqref{epsilon}, we arrive at the condition
\begin{equation}
\nab_i\nab_j\alpha=0\;.
\end{equation}
In other words, $(\mathscr{S},h)$ must possess a covariantly constant field $u_i$ such that $u_i=\nab_i\alpha$.  Excluding the trivial case $\alpha=\mbox{const}$, such a $u_i$ exists only if one can choose coordinates $(y,z^1,z^2)$ on $\mathscr{S}$ such that $u_i=\nab_i y$ and 
\begin{equation}
h=\dd y^2+G_{AB}\,\dd z^A\dd z^B\;,
\label{G}
\end{equation}
where $A$ and $B$ run from 1 to 2, and the coefficients $G_{AB}$ do not depend on $y$~\cite{stephani}.  The function $\alpha$ is then such that $\nab_i\alpha=\nab_i y$, so $\alpha=y+a$, where $a$ is a constant.

Since the $\beta^i$ are arbitrary functions of the spatial coordinates alone,  the space of ultrastatic transformations is infinite-dimensional~\cite{fn3} and contains the purely spatial, time-independent transformations $x^i\rightarrow\bar{x}^i$ as a subspace, generated by the vector $\beta:=\beta^i\partial_i$.  Similarly, the constant $a$ leads to time translations.  One can thus generate a proper ultrastatic transformation setting $\beta\equiv 0$ and $a\equiv 0$.  The previous result implies then that nontrivial proper ultrastatic transformations exist only if the metric can be written as
\begin{equation}
\g=-\dd t^2+\dd y^2+G_{AB}\,\dd z^A\dd z^B\;,
\label{gG}
\end{equation}
where the $G_{AB}$ depend only on the $z^A\,$s.  Therefore, an ultrastatic spacetime admits proper ultrastatic transformations iff it contains regions that are the product of a two-dimensional Minkowski spacetime and a two-dimensional space.  The proper ultrastatic transformations are just Lorentz boosts in these Minkowskian sections, with generator $y\,\partial_0+t\,\partial_y$ in the coordinates $(t,y,z^1,z^2)$.

The main conclusion can be reached in a more straightforward way by noting that, in order for nontrivial proper ultrastatic transformations to exist, the spacetime must possess a timelike Killing vector field $\zeta^a$ different from $\eta^a$, which is also hypersurface-orthogonal and has unit norm.  Such a vector field can always be written as $\zeta^a=\gamma\left(\eta^a+v\tau^a\right)$, where $\tau^a$ is a spacelike unit vector field orthogonal to $\eta^a$, $v$ is a function on $\mathscr{M}$ taking values in the interval $\left[0,1\right[\subset\mathbb{R}$, and $\gamma=\left(1-v^2\right)^{-1/2}$ because of normalisation.  Since $\nab_a\zeta_b=\nab_a\eta_b=0$, one finds that $\eta_b\nab_a\gamma+ \nab_a\left(\gamma v\tau_b\right)=0$.  Contracting with $\eta^b$ and using the properties $\eta^a\tau_a=0$ and $\nab_a\eta_b=0$, it follows that $\gamma$ (that is, $v$) must be a constant, hence that $\nab_a\tau_b=0$. Thus, nontrivial proper ultrastatic transformations exist iff, in addition to the timelike Killing vector field $\eta^a$, there is also a spacelike Killing vector field $\tau^a$ which has unit norm and is hypersurface-orthogonal~\cite{fn4}.  From this it follows that a chart can be found in which the metric tensor has the form~\eqref{gG}, and that a proper ultrastatic coordinate transformation is just a one-dimensional Lorentz boost in the direction of $\tau^a$.


\section{Conformal ultrastatic transformations}
\label{sec4}
\setcounter{equation}{0}

In an ultrastatic spacetime, let us suppose that there exists a coordinate transformation $x^a\to \bar{x}^a$ such that
\begin{equation}
-\dd t^2+\g_{ij}\,\dd x^i\dd x^j=\ee^{2\Phi}\left(-\dd \bar{t}^2+ \tilde{\g}_{ij}\,\dd\bar{x}^i\dd\bar{x}^j\right)\;,
\label{confcoord}
\end{equation}
where $\Phi$ is a regular non-constant function, and $\partial\tilde{\g}_{ij}/\partial\bar{t}=0$. In this case we shall speak of a {\em conformal ultrastatic transformation\/}.

That transformations of this type may indeed exist, can be realised by considering the particular case in which $\g_{ij}$ and $\tilde{\g}_{ij}$ are both flat metrics. In this case, it is well known that equation~\eqref{confcoord} holds for nontrivial $\Phi$'s, when $x^a\to \bar{x}^a$ is a special conformal coordinate transformation~\cite{conf,cft}.  As another very simple example, consider the transformations $t=:\ee^{x'}\sinh t'$, $x=:\ee^{x'}\cosh t'$ and $t=:\ee^{t''}\cosh x''$, $x=:\ee^{t''}\sinh x''$ in appropriate regions of two-dimensional Minkowski spacetime.  These lead, respectively, to 
\begin{equation}
-\dd t^2+\dd x^2=\ee^{2\,x'}\left(-\dd t'^2+\dd x'^2\right)
\label{wedge}
\end{equation}
(Rindler's metric~\cite{rindler}), and 
\begin{equation}
-\dd t^2+\dd x^2=\ee^{2\,t''}\left(-\dd t''^2+\dd x''^2\right)
\label{milne}
\end{equation}
(two-dimensional Milne universe~\cite{rindler}).  Hence, they are conformal ultrastatic transformations corresponding to $\Phi=x'=\ln\left(x^2-t^2\right)^{1/2}$ and $\Phi=t''=\ln\left(t^2-x^2\right)^{1/2}$, respectively~\cite{fn5}.

Equation~\eqref{confcoord} can be rewritten as
\begin{equation}
\ee^{2\Phi}\tilde{\g}_{ab}(\bar{x})=\bar{\g}_{ab}(\bar{x})=\g_{ab}(x)+\varepsilon\,\frac{\partial\g_{ab}(x)}{\partial x^c}\,\xi^c(x)-\varepsilon\,\pounds_\xi\,\g_{ab}(x)+{\cal O}(\varepsilon^2)\;,
\label{confcoordexp}
\end{equation}
where equation~\eqref{ggbar} has been used and $\g_{ab}(\bar{x})$ has been expanded around $x$.  The $00$ component of~\eqref{confcoordexp} becomes, remembering the conditions $\tilde{\g}_{00}(\bar{x})=\g_{00}(x)=-1$, 
\begin{equation}
\Phi=-\varepsilon\,\partial\alpha/\partial t+{\cal O}(\varepsilon^2)\;,
\label{00}
\end{equation}
where $\alpha:=\xi^0$.  The components $0i$ give again $\partial\xi_i/\partial t=\partial\alpha/\partial x^i$, so
\begin{equation}
\xi^i=\g^{ij}\,\partial_j\int\dd t\,\alpha+\beta^i\;,
\label{epsilon-conf}
\end{equation}
where the $\beta^i$ do not depend on $t$.  Finally, the $ij$ components, together with equation~\eqref{ttbar} and the conditions $\partial\g_{ij}(x)/\partial t=\partial\tilde{\g}_{ij}(\bar{x})/\partial\bar{t}=0$, give
\begin{equation}
\nab_i\nab_j\alpha=\frac{\partial^2\alpha}{\partial t^2}\,\g_{ij}\;.
\label{ij}
\end{equation}
If $\alpha$ does not depend on $t$, we recover the case discussed in section~\ref{sec3}.  Consistently, equation~\eqref{00} implies $\Phi={\cal O}(\varepsilon^2)$.  If, on the other hand, $\alpha$ does not depend on the spatial coordinates, then equation~\eqref{ij} implies $\partial^2\alpha/\partial t^2=0$, so $\alpha$ is a linear function of $t$ and $\Phi$ is a constant.  Interesting possibilities arise only if $\alpha$ depends on both $t$ and at least one of the spatial coordinates.  

In the particular case of Minkowski spacetime, equation~\eqref{ij} can be easily solved in Lorentzian coordinates and gives
\begin{equation}
\alpha=A\left(t\,\boldsymbol{x}^2+t^3/3\right)+B\left(t^2+\boldsymbol{x}^2\right)+\boldsymbol{C}\!\cdot\!\boldsymbol{x}\, t+\boldsymbol{D}\!\cdot\!\boldsymbol{x}+E t+K\;,
\label{sct}
\end{equation}
where $\boldsymbol{x}$ denotes the Cartesian vector with components $x^1$, $x^2$, $x^3$, and $A$, $B$, $\boldsymbol{C}$, $\boldsymbol{D}$, $E$, $K$ are constants.  Equation~\eqref{epsilon-conf} then yields
\begin{equation}
\xi^i=At^2 x^i+2Bx^i t+\frac{1}{2}\,C^i t^2+D^i t+\beta^i\;.
\end{equation}
One immediately recognizes in the quadratic terms of these expressions the generators of the special conformal transformations~\cite{conf,cft}.

Coming back to the general case, rewriting equation~\eqref{ij} in the equivalent form 
\begin{equation}
\nab_i\nab_j\alpha+\nab_j\nab_i\alpha=2\,\frac{\partial^2\alpha}{\partial t^2}\,\g_{ij}\;,
\label{ckvf}
\end{equation}
we see that there must be a conformal Killing field $u^i$ for $(\mathscr{S},h)$, such that $\g_{ij}u^j=\nab_i\alpha$ and  
\begin{equation}
\nab_i\, u^i=3\,\frac{\partial^2\alpha}{\partial t^2}\;.
\label{divu}
\end{equation}
Given an ultrastatic spacetime, the steps for finding the generators of its conformal ultrastatic transformations are then the following ones: (1) find the conformal Killing vector fields $u^i$ of the spatial metric $h$; (2) among these fields, determine those for which $u_i=\g_{ij}u^j$ is a gradient, so one can define $\alpha$ such that $\nab_i\alpha=u_i$; (3) integrate the one-form $u_i\dd x^i$ to find $\alpha$, using also equation~\eqref{divu}.  As an example, consider again Minkowski spacetime.  The vector field that, in Cartesian coordinates, has components 
\begin{equation}
u^i=2\lambda(t)\,x^i+\mu^i(t)\;, 
\label{confKM}
\end{equation}
where $\lambda$ and $\boldsymbol{\mu}$ are arbitrary functions and $t$ is treated as a parameter, is a conformal Killing vector field of the (Euclidean) spatial metric, and $u_i$ is the spatial gradient of 
\begin{equation}
\alpha=\lambda(t)\,\boldsymbol{x}^2+\boldsymbol{\mu}(t)\!\cdot\!\boldsymbol{x}+f(t)\;,
\label{alphaM}
\end{equation}
where $f$ is an arbitrary function.  Replacing the expression~\eqref{alphaM} into equation~\eqref{divu} we find $\lambda(t)=At+B$, $\boldsymbol{\mu}(t)=\boldsymbol{C}t+\boldsymbol{D}$, $f(t)=At^3/3+Bt^2+Et+K$, where $A$, $B$, $\boldsymbol{C}$, $\boldsymbol{D}$,  $E$, $K$ are constants.  The resulting $\alpha$ coincides with the one in equation~\eqref{sct}.  (Note that one cannot use the generator of a special conformal transformation for the same purpose, because the associated one-form is not closed.)

Let us now remind~\cite{ottewill} that if $u^i$ is a conformal Killing vector field of $(\mathscr{S},h)$, then it is a Killing vector field of $(\mathscr{S},\Omega^2h)$, for 
\begin{equation}
\Omega:=\left(\g_{ij}\,u^i u^j\right)^{-1/2}\;.
\end{equation}
Since, in addition, $u^i$ is hypersurface-orthogonal in our case, it follows that there must exist coordinates $(y,z^1,z^2)$ on $\mathscr{S}$, such that $u^i=\delta^i_y$ and 
\begin{equation}
\Omega^2\,h=H(t,z)\, \dd y^2+G_{AB}(t,z)\,\dd z^A\dd z^B\;,
\label{G'}
\end{equation}
where $A$ and $B$ run from 1 to 2.  Note that since $\alpha$ depends on $t$ in general, and $\g_{ij}u^j=\nab_i\alpha$, the coordinate transformation on $\mathscr{S}$ that leads to $(y,z^1,z^2)$ may contain $t$ as a parameter; this explains the possible $t$-dependence of the new metric coefficients.  Since $\Omega^2\g_{ij}u^iu^j=1$, we find $H=1$, so 
\begin{equation}
h=\Omega^{-2}\left(\dd y^2+G_{AB}\,\dd z^A\dd z^B\right)\;.
\label{spacemetric}
\end{equation}
Hence, an ultrastatic spacetime admits proper conformal ultrastatic transformations iff it contains regions where the spatial sections are  conformal to a product space.  

{From} the equality $u_i\dd x^i=u_y\dd y=\Omega^{-2}\dd y$, it follows that the coordinate $y$ is defined by the relation
\begin{equation}
\dd y=\frac{u_i\dd x^i}{\Omega^{-2}}=\frac{\g_{ij}u^j\dd x^i}{\g_{kl}u^k u^l}\;.
\label{dy}
\end{equation}
Moreover, $\nab_i\alpha=u_i$ implies that $\partial_y\alpha=\Omega^{-2}$ and $\partial_A\alpha=0$, so $\alpha$ and $\Omega$ do not depend on the coordinates $z^1$ and $z^2$, and
\begin{equation}
\alpha=\int\dd y\;\Omega^{-2}+\phi(t)\;,
\label{alpha}
\end{equation}
where $\phi$ is an unspecified function.  For example, the conformal Killing vector field~\eqref{confKM} in Minkowski spacetime gives 
\begin{equation}
\Omega^{-2}=4\lambda(t)^2\boldsymbol{x}^2+4\lambda(t)\,\boldsymbol{\mu}(t)\!\cdot\!\boldsymbol{x}+\boldsymbol{\mu}(t)^2\;,
\end{equation}
and it is easy to check that $\dd\Omega^{-2}=4\lambda(t)\,u_i\dd x^i$, so $\Omega^{-2}={\rm e}^{4\lambda(t)\,y}$ by equation~\eqref{dy}.  Equation~\eqref{alpha} then yields
\begin{equation}
\alpha=\frac{{\rm e}^{4\lambda(t)\,y}}{4\lambda(t)}+\phi(t)=\lambda(t)\,\boldsymbol{x}^2+\boldsymbol{\mu}(t)\!\cdot\!\boldsymbol{x}+\frac{\boldsymbol{\mu}(t)^2}{4\lambda(t)}+\phi(t)\;.
\label{alphanew}
\end{equation}
Replacing this into equation~\eqref{divu} we find the previous expressions for $\lambda(t)$ and $\boldsymbol{\mu}(t)$, and 
\begin{equation}
\phi(t)=\frac{1}{3}\,A t^3+Bt^2-\frac{\boldsymbol{\mu}(t)^2}{4\lambda(t)}+E t+K\;,
\end{equation}
which leads again to the expression~\eqref{sct} for $\alpha$.


\section{Conclusions}
\label{sec5}
\setcounter{equation}{0}

I have described the general properties of ultrastatic spacetimes.  Excluding the very special cases of spacetimes containing regions that are products with a Minkowskian factor, I have shown that canonical coordinates are unique, up to time translations and purely spatial coordinate  transformations.  I have also shown that an ultrastatic spacetime is non-trivially conformal to other ultrastatic spacetimes iff it contains regions whose spatial sections are conformal to a product (equivalently, iff it admits conformal Killing vector fields whose associated one-form is exact).  

In the formalism discussed in~[3], the gravitational potential in a conformally static spacetime $(\mathscr{M},\g)$ is associated with the conformal factor that links $\g$ to an ultrastatic metric.  Such a metric is also employed in many calculations~\cite{???,optmet}.  If proper conformal ultrastatic transformations exist, the ultrastatic metric is not unique, which is a potentially dangerous circumstance for these approaches.  The results of section~\ref{sec4} establish when this can be a concern.


\section*{Acknowledgements}

I am grateful to an anonymous referee for comments that stimulated improvements in the presentation.



\begin{thebibliography}{99}
\bibitem{fn1}
Latin letters $a$, $b$, ... and $i$, $j$, ... from the beginning and the middle of the alphabet denote tensorial components in some chart, and run from 0 to 3 and from 1 to 3, respectively.  I work in units in which $c=1$, and choose +2 as signature of the metric.
\bibitem{discrete}
I focus on transformations that admit generators, so in particular I am not dealing with discrete transformations.
\bibitem{fn4}
This result could also be obtained looking for the $\tau^a$ such that the infinitesimal transformation $\eta^a\rightarrow\zeta^a:=\eta^a+v\tau^a$ gives a new field $\zeta^a$ which, to first order in $v$, is also normalised to $-1$ and covariantly constant. One easily finds that $\eta^a\tau_a=0$ and $\nab_b\tau^a=0$.
\bibitem{fn5}
The two-dimensional metrics are both flat iff their scalar curvatures (containing the same information as the full Riemann tensor) are $R=\widetilde{R}=0$. Equation~(D.9) of reference~[8] gives then $\g^{ab}\nab_a\nab_b\Phi=0$, which implies $\Phi(t,x)=F(x-t)+G(x+t)$, where $(t,x)$ are coordinates that are Lorentzian with respect to $\g$, and $F$, $G$ are arbitrary $C^2$ functions.  Consequently, the conformal group --- hence, the space of conformal ultrastatic transformations --- is in this case infinite-dimensonal (see reference~[14]).  For $G(x+t)=\left(x+t\right)^{1/2}$, the choices $F(x-t)=\ln\left(x-t\right)^{1/2}$ and $F(x-t)=\ln\left(t-x\right)^{1/2}$ lead, respectively, to equations~\eqref{wedge} and~\eqref{milne}.
\end{thebibliography}
\end{document}